\documentclass[
showkeys,12pt,
preprint,preprintnumbers,nofootinbib,
groupedaddress,superscriptaddress,amsmath,amssymb]{revtex4}
\usepackage{graphicx}
\usepackage{dcolumn}
\usepackage{bm}
\usepackage{amssymb}
\usepackage{amsmath}
\usepackage{epsfig}    
\usepackage{color}
\usepackage{slashed}
\usepackage{hhline}

\def\be{\begin{equation}}
\def\ee{\end{equation}}
\newcommand{\bea}{\begin{eqnarray}}
\newcommand{\eea}{\end{eqnarray}}
\newcommand{\nn}{\nonumber}

\numberwithin{equation}{section}

\begin{document}

\title{
Radiative Seesaw in Minimal 3-3-1 Model
}
\preprint{KIAS-P15017}
\author{Hiroshi Okada}
\email{hokada@kias.re.kr}
\affiliation{School of Physics, KIAS, Seoul 130-722, Korea}
\author{Nobuchika Okada}
\email{okadan@ua.edu}
\affiliation{Department of Physics and Astronomy, University of Alabama, Tuscaloosa, AL35487, USA}
\author{Yuta Orikasa}
\email{orikasa@kias.re.kr}
\affiliation{School of Physics, KIAS, Seoul 130-722, Korea}
\affiliation{Department of Physics and Astronomy, Seoul National University, Seoul 151-742, Korea}

\date{\today}

\begin{abstract}
We study the neutrino sector in a minimal $SU(3)_L\times U(1)_X$ model, 
  in which its mass is generated at one-loop level with the charged lepton mass, and hence 
  there exists a strong correlation between the charged-lepton mass and the neutrino mass. 
We identify the parameter region of this model to satisfy the current neutrino oscillation data 
  as well as the constraints on lepton flavor violating processes. 
We also discuss a possibility to explain the muon anomalous magnetic moment.    
\end{abstract}
\maketitle
\newpage

\section{Introduction}
The flavor problem can be one of the biggest challenging issues to be resolved, since the standard model (SM) does not have 
any theoretical sources (especially) why the number of family for each fermion sector (quarks and leptons) is three.
One of the reasonable interpretations is to extend the gauge sector $SU(2)_L\times U(1)_Y$ to $SU(3)_L\times U(1)_X$, 
so called 3-3-1 model, in which the origin of three family is coming from the number of $SU(3)$ color of quarks (that has three) due to 
the gauge anomaly cancellation~\cite{Singer:1980sw, Valle:1983dk}.
Because of larger  gauge group comparing to the SM one, there are several variations of models extending the Higgs sector~
\cite{Pisano:1991ee, Frampton:1992wt, Foot:1992rh, Hoang:1995vq, Frampton:1993wu} and revisited models to reanalyze with current experimental data~\cite{Boucenna:2014ela, Kelso:2014qka, daSilva:2014qba, Hernandez:2015tna, Hernandez:2015cra, Martinez:2014lta, Dong:2014esa, Phong:2014yca, Montalvo:2014gka, DeConto:2014fza, Phong:2014ofa, Montero:2014uya, Dong:2014bha, Pires:2014xsa, Dong:2015rka, Benavides:2015afa, Salazar:2015gxa, Queiroz:2010rj, Boucenna:2014dia, Boucenna:2015zwa, Long:2015gca, Binh:2015cba}.

On the other hand, explaining the current neutrino oscillation data and dark matter (DM) candidate might be done by physics beyond SM.
It is known that in recent radiative seesaw models they can be only explained simultaneously but also correlated each other.
It means that neutrinos do not directly interact with the SM Higgs field but with a DM candidate. As a result,  minuscule neutrino masses can be 
naturally realized. This is because a vast literature has recently arisen along thought of  this subject~\cite{Zee, Zee-Babu, KNT, Ma, Kanemura:2013qva, Aoki:2008av, Kanemura:2014rpa, Kanemura:2011mw, Kajiyama:2013zla, Gustafsson:2012vj, Hatanaka:2014tba, Jin:2015cla}.

In this paper, we combine the 3-3-1 model and radiative seesaw model based on a minimal model in Ref.~\cite{Foot:1992rh}, in which we do not impose any additional discrete symmetry. Then we can generate the neutrino mass at one-loop level.
Moreover since our model has a strong correlation between the charged lepton sector and the neutrino sector due to the same origin of these masses, 
it may be worth analyzing the neutrino oscillation data as well as lepton flavor violating processes and so on.

This paper is organized as follows.
In Sec.~II, we show our model including Higgs potential.
In Sec.~III, we analyze lepton sector and show how to correlate the charged-lepton masses and neutrino masses.
Then we also show to compute  lepton flavor violating processes and muon anomalous magnetic moment.
In Sec.~IV, we perform parameter scan to identify allowed parameter regions. 
We conclude in Sec.~V.


 \begin{widetext}
\begin{center} 
\begin{table}[b]
\begin{tabular}{|c||c||c|c|c|c|}\hline  
&\multicolumn{1}{c||}{Lepton Fields} & \multicolumn{4}{c|}{Scalar Fields} \\\hline\hline
& ~$L_L=(\nu_L,e_L,e^c_R)$~ & ~$S$~ & ~$\eta$~ & ~$\rho$~  & ~$\chi$ \\\hline 
$SU(3)_L$ & $\bm{3}$ & $\bm{6}$& $\bm{3}$ & $\bm{3}$ & $\bm{3}$\\\hline 
$U(1)_X$ & $0$ & $0$ & $0$ & $1$  & $-1$   \\\hline
\end{tabular}
\caption{Contents of lepton and scalar fields
and their charge assignment under $SU(3)_L\times U(1)_X$, where the index of the generation are abbreviated.}
\label{tab:1}
\end{table}
\end{center}
\end{widetext}

\section{ Model setup}
We discuss a possibility of a one-loop induced radiative seesaw model in the context of 3-3-1 model in \cite{Foot:1992rh}. 
The particle contents are shown in Tab.~\ref{tab:1}. 
We introduce a gauge triplet fermion $L_L=(\nu_L,e_L,e^c_R)$ with $U(1)_X$=0. {For new bosons, we introduce $SU(3)_L$ sextet scalars $S$ with $U(1)_X$=0,
 $SU(3)_L$ triplet scalars ($\eta,\rho,\chi$) with $U(1)_X=(0,1,-1)$, respectively.
The renormalizable Lagrangian for Lepton Yukawa sector, and the scalar potential
under these assignments are given by
\begin{align}
\mathcal{L}_{Y}
&=
y_{\ell_1}\sum_{i,j,k=1-3} \bar L_{Li} (L_{L})^c_j \eta^*_k\epsilon^{ijk} + y_{\ell_2}{\rm Tr}[ \bar L_{L} S (L_{L})^c ] + \rm{h.c.} 
\label{eq:lepton-mass}
\\ 
\mathcal{V}
&= 
 m_\eta^2 |\eta|^2 + m_{\rho}^2 |\rho|^2  + m_{\chi}^2 |\chi|^2  +  m_{S}^2 {\rm Tr}[|S|^2]
 \nn\\
&
  + \lambda_1 |\eta|^{4}   + \lambda_2 |\rho|^{4}   + \lambda_3 |\chi|^{4}   + \lambda_4 [{\rm Tr}[|S|]]^{4}   + \lambda_5 {\rm Tr}[|S|^4] 
  + \lambda_6 |\eta|^2|\rho|^2 
  + \lambda_7 |\eta|^2|\chi|^2 
  + \lambda_8 |\chi|^2|\rho|^2 
  \nn\\
&   
  + \lambda_9 |\eta|^2{\rm Tr}[|S|^2] 
 + \lambda_{10} |\rho|^2{\rm Tr}[|S|^2]  + \lambda_{11} |\chi|^2{\rm Tr}[|S|^2] 
+
 \lambda_{12} |\rho^\dag\eta|^2  + \lambda_{13} |\chi^\dag\eta|^2  + \lambda_{14} |\rho^\dag\chi|^2 
   \nn\\
&  
+(m_1\sum_{i,j,k=1-3}\epsilon^{ijk}\eta_i\rho_j\chi_k+{\rm h.c.} ) + (m_2 \rho^T S^\dag\chi+{\rm h.c.} ) 
+ (m_3 \eta^T S\eta+{\rm h.c.} )
\nn\\&
+(m_4\sum_{i,j,k}^{1-3} \sum_{l,m,n}^{1-3} \epsilon^{ijk} \epsilon^{lmn} S_{il} S_{jm} S_{kn}+{\rm h.c.} ) 
+(f_5(\eta^\dag\rho)(\eta^\dag\chi)+{\rm h.c.})
+(f_6\sum_{i,j,k}^{1-3}  \epsilon^{ijk} S_{li} \rho_{j} \chi_k \eta^*_{l}+{\rm h.c.} ) 
\nn\\&
+f_7{\rm Tr}[S^\dag S \eta^*\eta^T]+f_8{\rm Tr}[S^\dag S \rho^*\rho^T]+f_9{\rm Tr}[S^\dag S \chi^*\chi^T]
+(f_{10 }\sum_{i,j,k}^{1-3} \sum_{l,m,n}^{1-3} \epsilon^{ijk} \epsilon^{lmn} S_{il} S_{jm} \eta_{k}\eta_{n}+{\rm h.c.} ) 
,
\label{HP}
\end{align}
where $y_{\ell_1}$ is an anti-symmetric 3 by 3 matrix and $y_{\ell_2}$ is a symmetric one. 
Here the scalar fields can be parameterized as 
\begin{align}
S &=\left[
\begin{array}{ccc}
\sigma_1^0 & h_2^- & h_1^+\\
h_2^- & H_1^{--} & \sigma_2^0\\
h_1^+ & \sigma_2^0 & H_2^{++}
\end{array}\right],\quad
\eta =\left[
\begin{array}{c}
\eta^0\\\eta^-_1\\ \eta^+_2
\end{array}\right],\quad 
\rho =\left[
\begin{array}{c}
\rho^+\\ \rho^0\\ \eta^{++}
\end{array}\right],\quad
\chi =\left[
\begin{array}{c}
\chi^-\\ \chi^{--}\\ \chi^{0}
\end{array}\right],  
\sigma_1^0=\frac{\sigma_{1R}+i \sigma_{1I}}{\sqrt2},\quad \\
\sigma_2^0&=\frac{v_\sigma+\sigma_{2R}+i \sigma_{2I}}{\sqrt2},\quad
\eta^0=\frac{v_\eta+\eta_{R}+i \eta_{I}}{\sqrt2},\quad
\rho^0=\frac{v_\rho+\rho_{R}+i \rho_{I}}{\sqrt2},\quad
\chi^0=\frac{v_\chi+\chi_{R}+i \chi_{I}}{\sqrt2}.
\label{component}
\end{align}
Here  $v  = \sqrt{v_\sigma^2+v_\eta^2+v_\rho^2}= 246$ GeV is the vacuum expectation value (vev), 
  and we have assumed that the vev of $\sigma_1^0$ is zero, which is unlike the original model discussed in~\cite{Foot:1992rh}. 
The scale of $v_\chi$, which breaks $SU(3)_L$ symmetry,  is assumed to be of ${\cal O}$(1) TeV,
{while the other three vevs ($v_\sigma,v_\rho,v_\eta$) are assumed to be ${\cal O}$(1-100) GeV.}

\section{Lepton sector}

\subsection{ Charged lepton sector}
The charged lepton masses can be generated by the terms $y_{\ell_1}\sum_{i,j,k=1-3} \bar L_{Li} (L_{L})^c_j \eta^*_k\epsilon^{ijk}$ and $y_{\ell_2}{\rm Tr}[ \bar L_{L} S (L_{L})^c ]$ after the electroweak symmetry breaking, as can be seen in Eq.~(\ref{eq:lepton-mass}). The resulting mass matrix can be written and diagonalized by
\begin{align}
&(\bar e_{L})_a (m_\ell)_{ab}  (e_{R})_b
\equiv
(\bar e_{L})_a \left[\frac{(y_{\ell_1}) v_\eta} {\sqrt2} + \frac{(y_{\ell_2}) v_\sigma} {\sqrt2} \right]_{ab}  (e_{R})_b=
(\bar e_{L})_a (V^\dag _{eL})_{ai} (m^{diag}_\ell)_{i}  (V _{eR})_{ib}  (e_{R})_{b},\label{eq:ele-form}\\
&\sum_{k=1}^3
 \left[\frac{(y_{\ell_1}) v_\eta} {\sqrt2} + \frac{(y_{\ell_2}) v_\sigma} {\sqrt2} \right]_{ak} 
  \left[\frac{(y_{\ell_1}) v_\eta} {\sqrt2} + \frac{(y_{\ell_2}) v_\sigma} {\sqrt2} \right]^*_{bk}
  = (V^\dag _{eL})_{ai} |m^{diag}_\ell|^2_{i}  (V _{eL})_{ib} , 
\end{align}
where $y_{\ell_1}$ is the ant-symmetric matrix and $y_{\ell_2}$  is the symmetric matrix. $ |m^{diag}_\ell|=(|m_e|,|m_\mu|,|m_\tau|)
=(0.5\ {\rm MeV}, 105.6\ {\rm MeV}, 1777\ {\rm MeV})$.
Each of $V_{eL}$ and $V_{eR}$ is the unitary  matrix to diagonalize the charged leptons. 
{
From Eq.~(\ref{eq:ele-form}) and the properties of anti-symmetric $y_{\ell_1}$ and of symmetric $y_{\ell_2}$, the charged lepton Yukawa couplings $y_{\ell_{1(2)}}$ can explicitly be rewritten in terms of the charged lepton mass matrix as
\begin{align}
(y_{\ell_{1}})_{ab} =\frac{(m_\ell)_{ab} + (m_\ell)_{ab}^T }{\sqrt2 v_\eta},\quad
(y_{\ell_{2}})_{ab} =\frac{(m_\ell)_{ab} -(m_\ell)_{ab}^T }{\sqrt2 v_\sigma}.
\end{align}  
Therefore, these Yukawa couplings can be taken as the output parameters, once the charged lepton masses and their mixings are 
fixed as inputs in the following numerical analysis.
}

\subsection{ Neutrino sector}
\begin{figure}[tbc]
\begin{center}
\includegraphics[scale=0.6]{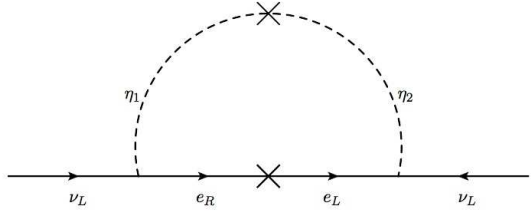}
\caption{
An example of one-loop diagrams contributing to neutrino mass. 
 }

\label{fig:loop}
\end{center}
\end{figure}

Through the following Yukawa interactions, 
\begin{eqnarray}
{\cal L} &\supset& 
y_{\ell_1}(\bar\nu_L e^c_L - \bar e_L \nu^c_L)\eta^-_2
+y_{\ell_1}(-\bar\nu_L e_R + \bar e^c_R \nu^c_L)\eta^+_1 
\nonumber \\
&+&
y_{\ell_2}(\bar\nu_L e^c_L + \bar e_L \nu^c_L) h^-_2
+y_{\ell_2}(\bar\nu_L e_R + \bar e^c_R \nu^c_L) h^+_1, 
\label{eq:laglep} 
\end{eqnarray}
the active neutrino mass matrix $m_\nu$ is induced at one-loop level 
 (see  Fig.~\ref{fig:loop} for an example diagram), which is given by 
\begin{align}
-({\cal M}_{\nu}^{\rm th} )_{ab}
&\approx
\frac{\left[y_{\ell_1} \left(m^\dag_\ell +m^{*}_\ell \right) y^T_{\ell_1}\right]_{ab}}{(4\pi)^2} 
\left[\frac{\delta m^{+2}_{\eta_1\eta_2}}{m_{\eta^+_1}^{2}-m_{\eta^+_2}^2}\right]{\ln\left[\frac{m_{\eta^+_1}^{2}}{m_{\eta^+_2}^{2}}\right]}\nn\\
&-
\frac{\left[y_{\ell_2} \left(m^\dag_\ell +m^{*}_\ell \right) y^T_{\ell_2}\right]_{ab}}{(4\pi)^2} 
\left[\frac{\delta m^{+2}_{h_1\eta_2}}{m_{h^+_1}^{2}-m_{h^+_2}^2}\right]{\ln\left[\frac{m_{h^+_1}^{2}}{m_{h^+_2}^{2}}\right]}
\nn\\
&-
\frac{\left(y_{\ell_1} m^{*}_\ell y^T_{\ell_2}+y_{\ell_2} m^{\dag}_\ell y^T_{\ell_1}\right)_{ab}}{2(4\pi)^2} 
\left[\frac{\delta m^{+2}_{\eta_2h_1}}{m_{\eta^+_2}^{2}-m_{h^+_1}^2}\right]{\ln\left[\frac{m_{\eta^+_2}^{2}}{m_{h^+_1}^{2}}\right]}\nn\\
&-
\frac{\left(y_{\ell_2} m^{*}_\ell y^T_{\ell_1}+y_{\ell_1} m^{\dag}_\ell y^T_{\ell_2}\right)_{ab}}{2(4\pi)^2} 
\left[\frac{\delta m^{+2}_{\eta_1h_2}}{m_{\eta^+_1}^{2}-m_{h^+_2}^2}\right]{\ln\left[\frac{m_{\eta^+_1}^{2}}{m_{h^+_2}^{2}}\right]}.
\label{eq:numass}
\end{align}
Note here that the mass insertion approximation has been used, that is, $(M_+)^2= (M_+)^2_{\rm diag}+(M_+)^2_{\rm off-diag}$ with $ (M_+)^2_{\rm diag}  \gg (M_+)^2_{\rm off-diag}\equiv \delta m^{+2}_{f_if'_j},\ (f_i,f'_j=\eta_i, h_j)$, where $M_+$ is the singly charged boson mass matrix, and 
{
\begin{eqnarray} 
&&
m^2_{h_1^+}\approx f_9 v^2_\chi,\quad m^2_{h_2^+}\approx \frac{m_2 v_\chi v_\rho}{v_\sigma},\quad
m^2_{\eta_1^+}\approx \frac{m_1 v_\chi v_\rho}{v_\eta},\quad m^2_{\eta_2^+}\approx {\lambda_{13} v_\chi^2 },
\nn\\
&& \delta m^{+2}_{\eta_1\eta_2}= \sqrt2 m_3 v_\eta +(f_5 v_\rho v_\chi)/2,  \; 
\delta m^{+2}_{h_1h_2}=3\sqrt2 m_4 v_\sigma ,  
\nonumber \\
&& 
\delta m^{+2}_{\eta_1 h_1} =(4 f_{10} + f_7) v_\sigma v_\eta/(2 \sqrt{2}),  \; 
\delta m^{+2}_{\eta_2 h_1} = m_3 v_\eta +(f_6 v_\rho v_\chi)/(2 \sqrt{2}),   
\nonumber \\
&&
\delta m^{+2}_{\eta_1 h_2} = m_3 v_\eta +(f_6 v_\rho v_\chi)/(2 \sqrt{2}),  \; 
\delta m^{+2}_{\eta_2 h_2} =(4 f_{10} + f_7) v_\sigma v_\eta/(2 \sqrt{2}). 
\end{eqnarray}
It suggests 
$m_1,m_2\gg m_3, m_4$ and $f_9,\lambda_{13}\gg f_5,f_6,f_7, f_{10}$.
}

Neutrino mass matrix can be diagonalized by a unitary matrix $U_\nu$ as $m_\nu=U_\nu m_\nu^{diag} U^T_\nu$, 
and hence the MNS matrix $U_{\rm MNS}$ is defined as 
\begin{align}
U_{\rm MNS}=V^\dag_{eL} U_\nu.
\end{align}

\subsection{Lepton Flavor Violations}
 Here we discuss the lepton flavor violating processes in our model, 
 which can be induced through the mixings among singly charged bosons in Eq.~(\ref{eq:laglep}). 
The branching ratio of the lepton flavor violating decays are given by 
\begin{align}
{\rm BR}(\ell_i\to\ell_j \gamma)&=\frac{12 \pi^2}{m_i^2 {\rm G_F}^2}\left[ | A_{Lji}|^2 + | A_{Rji}|^2 \right], 
\end{align}
where 
\begin{align}
A_{Lji}&=
m_j\left[G^{(ii)}_L(\eta_2,h_2)+G^{(iii+v)}_L(\eta_2,h_2)\right]_{ji}+m_i\left[G^{(i)}_L(\eta_1,h_1)+G^{(iv+vi)}_L(\eta_1,h_1)\right]_{ji},\\
A_{Rji}&=
m_j\left[G^{(i)}_R(\eta_1,h_1)+G^{(iv+vi)}_R(\eta_1,h_1)\right]_{ji}+m_i\left[G^{(ii)}_R(\eta_2,h_2)+G^{(iii+v)}_R(\eta_2,h_2)\right]_{ji},
\end{align}
\begin{align}
&G^{(i)}_{L(R)}(\eta_1,h_1)\approx
\frac{\delta m^{+2}_{\eta_1h_1}[(y_{\ell_2}^\dag y_{\ell_1})'+(y_{\ell_1}^\dag y_{\ell_2})']}{12(4\pi)^2m_{\eta_1}^2 m_{h_1}^2},\\
&G^{(ii)}_{L(R)}(\eta_2,h_2)\approx
\frac{\delta m^{+2}_{\eta_2h_2}[(y_{\ell_1} y_{\ell_2}^\dag)'+(y_{\ell_2} y_{\ell_1}^\dag)']}{12(4\pi)^2m_{\eta_2}^2 m_{h_2}^2}
,\\
&G^{(iii+v)}_{L(R)}(\eta_1,h_1)\approx
G^{(iv+vi)}_{L(R)}(\eta_2,h_2)\approx
\frac{1}{12(4\pi)^2}\left[\frac{(y_{\ell_1} y_{\ell_1}^\dag)'} {m_{\eta_2}^2 }+
\frac{(y_{\ell_2} y_{\ell_2}^\dag)'} {m_{h_2}^2 }\right],
\end{align}
%
%
with $(y_{\ell_a}^\dag y_{\ell_b})'\equiv V_{eR} y_{\ell_a}^\dag y_{\ell_b} V_{eR}^\dag$, and $(y_{\ell_a} y_{\ell_b}^\dag)'\equiv V_{eL} y_{\ell_a} y_{\ell_b}^\dag V_{eL}^\dag$, with $(a,b)=1,2$.
The upper indices of $G$, respectively, represent a difference among  the singly charged bosons exchanging processes.
Therefore, $(i)$ is the $\eta_1^+$ and $h_1^+$ mixing process, $(ii)$ is the $\eta_2^+$ and $h_2^+$ mixing one, $(iii)$ is the $\eta_1^+$ no mixing one, $(iv)$ is the  $\eta_2^+$ no mixing one, $(v)$ is the  $h_1^+$ no mixing one, and $(vi)$ is the  $h_2^+$ no mixing one.

{
 There are box types of LFV precesses  $\ell_a^-\to\ell_b^-\ell_c^+\ell_d^-$ through $y_{\ell_{1,2}}$ in general, however we expect such processes are negligibly small when our mass insertion approximation is reliable.
The penguin types of these LFV processes are also negligible, comparing to the $\ell_i \to\ell_j\gamma$ one-loop induced processes~\cite{Toma:2013zsa}.}
%

\subsection{Muon Anomalous Magnetic Moment}
The muon anomalous magnetic moment, so-called the muon $g-2$, has been 
measured at Brookhaven National Laboratory. 
The current average of the experimental results is given by~\cite{bennett}
\begin{align}
a^{\rm exp}_{\mu}=11 659 208.0(6.3)\times 10^{-10}. \label{obs}
\end{align}
It has been known that there is a discrepancy from the SM prediction by $3.2\sigma$~\cite{discrepancy1} 
to $4.1\sigma$~\cite{discrepancy2}:
\begin{align}
\Delta a_{\mu}=a^{\rm exp}_{\mu}-a^{\rm SM}_{\mu}=(29.0 \pm
9.0\ {\rm to}\ 33.5 \pm
8.2)\times 10^{-10}. \label{g-2_dev}
\end{align}
The contribution by the singly charged bosons is evaluated as 
\begin{align}
\Delta a_\mu
=
-\frac{m_\mu^2}{2}
\left[ G^{(i)}_L(\eta_1,h_1) + G^{(i)}_R(\eta_1,h_1) + G^{(ii)}_L(\eta_2,h_2) + G^{(ii)}_R(\eta_2,h_2)\right.\nn\\
\left.
+2G^{(iii+v)}_{L}(\eta_1,h_1) + 2 G^{(iv+vi)}_{L}(\eta_2,h_2)
\right]_{i=j=\mu}.
\end{align}

\section{Numerical Analysis}
For simplicity, we assume 
{the hierarchical neutrino mass spectrum and set the lightest eigenvalue to be zero.}
Then we parametrize  the neutrino mass matrix as 
\begin{align}
\mathcal{M}_\nu^{\rm exp} & = U_{\text{MNS}}\, {\rm diag}(m_{\nu_1},m_{\nu_2},m_{\nu_3})\,U_{\text{MNS}}^T,
 \label{mns0}
\end{align}
with the (real) standard  form of the MNS matrix, 
\begin{align}
U_{\text{MNS}}
 \equiv \left[ \begin{array}{ccc}
1 & 0 & 0 \\
0 & c_{23}^{} & s_{23}^{}  \\
0 & -s_{23}^{} & c_{23}^{}
\end{array}\right]
\left[ \begin{array}{ccc}
c_{13}^{} & 0 & s_{13}^{}
 \\
0 & 1 & 0  \\
-s_{13}^{}
& 0 & c_{13}^{} 
\end{array}\right]
\left[ \begin{array}{ccc}
c_{12}^{} & s_{12}^{} & 0 \\
-s_{12}^{} & c_{12}^{} & 0  \\
0 & 0 & 1\end{array}\right],
 \label{mns}
\end{align}
with $s_{ij}=\sin\theta_{ij}$ and $c_{ij}=\cos\theta_{ij}$. 
Then, one finds  a relation between $\mathcal{M}_\nu^{\rm th}$ and $\mathcal{M}_\nu^{\rm exp}$ given by 
\begin{align}
\mathcal{M}_\nu^{\rm exp}= V_{eL}^\dag \mathcal{M}_\nu^{\rm th} V_{eL},
\end{align}
where we also define $V_{eL(R)}$ as the standard parametrization as analogy to the MNS matrix, 
\begin{align}
V_{eL(R)}
 \equiv \left[ \begin{array}{ccc}
1 & 0 & 0 \\
0 & c_{eL(R)23}^{} & s_{eL(R)23}^{}  \\
0 & -s_{eL(R)23}^{} & c_{eL(R)23}^{}
\end{array}\right]
\left[ \begin{array}{ccc}
c_{eL(R)13}^{} & 0 & s_{eL(R)13}^{}
 \\
0 & 1 & 0  \\
-s_{eL(R)13}^{}
& 0 & c_{13}^{eL(R)} 
\end{array}\right]
\left[ \begin{array}{ccc}
c_{eL(R)12}^{} & s_{eL(R)12}^{} & 0 \\
-s_{eL(R)12}^{} & c_{eL(R)12}^{} & 0  \\
0 & 0 & 1\end{array}\right],
 \label{mns}
\end{align}
with $s_{eL(R)ij}^{}=\sin\theta_{eL(R)ij}$ and $c_{eL(R)ij}^{}=\cos^{}\theta_{eL(R)ij}$.

We perform parameter scan to reproduce the following neutrino oscillation data at 95\% confidence level~\cite{pdf} in the next subsection;
%
\begin{eqnarray}
&& 0.2911 \leq s_{12}^2 \leq 0.3161, \; 
 0.5262 \leq s_{23}^2 \leq 0.5485, \;
 0.0223 \leq s_{13}^2 \leq 0.0246,  
  \\
&& 
  \ |m_{\nu_3}^2- m_{\nu_2}^2| =(2.44\pm0.06) \times10^{-3} \ {\rm eV}^2,  \; 
  \ m_{\nu_2}^2- m_{\nu_1}^2 =(7.53\pm0.18) \times10^{-5} \ {\rm eV}^2. \nn
  \end{eqnarray}
%
%
%
%
We also impose the current experimental bounds 
on the lepton flavor violating processes~\cite{meg, Beringer:1900zz}:  
\begin{align}
{\rm BR}(\mu\to e\gamma)<5.7\times 10^{-13},\quad 
{\rm BR}(\tau\to \mu\gamma)<4.4\times 10^{-8},\quad 
{\rm BR}(\tau\to e\gamma)<3.3\times 10^{-8}.
\end{align}

\subsection{Normal ordering}
In case of the normal ordering, {$(m_{\nu_1},m_{\nu_2},m_{\nu_3})=(0,m_{\nu_2},m_{\nu_3})$},
 the input parameters vary in the following ranges, 
\begin{align}
&
1\ {\rm GeV}\le (v_\eta,v_\sigma)\le 100\ {\rm GeV},
\ 
0.1\ {\rm GeV^2}\le (\delta m^{+2}_{\eta_1 h_1},\delta m^{+2}_{\eta_2 h_1}) \le 10\ {\rm GeV^2},
\nn\\
&
100\ {\rm GeV}\le m_{\eta_{1,2}^+}\le1000\ {\rm GeV},\quad
-1\le (s_{eL(R)ij}^{}, {s}_{eL(R)ij}^{})\le 1. \label{eq:input}
\end{align}
{The mass parameters 
$m_{h_{1,2}^+}$, $\delta m^{+2}_{\eta_1 \eta_2}$, and $\delta m^{+2}_{h_1 h_2}$ can be written in terms of these input parameters
{from Eqs.~(\ref{eq:numass}) and (\ref{mns0}).
And} their
mass scales are found to be $ m_{h_{1,2}^+}\approx {\cal O}$(100) GeV, and $\delta m^{+2}_{h_1 h_2} \approx \delta m^{+2}_{\eta_1 \eta_2}\approx  {\cal O}(1)\ {\rm GeV^2}$.}
{Totally we have {\it twelve} input parameters shown in Eq.~(\ref{eq:input}) to reproduce the three lepton masses (3 charged lepton masses and 2 neutrino mass differences) and three mixings (3 MNS mixings). 
Once these input parameters are determined, all the physical values are uniquely fixed as discussed in the previous subsections.
Then we can compare the observable or upper constraints from each of the experimental result such as neutrino oscillation data and LFVs.}

Here we show some representative figures of our results which simultaneously satisfy the neutrino oscillation data 
and the constraints from the LFV processes. Here we have examined $10^7$ sampling points to search for our allowed parameters. 
The left panel of Fig.~\ref{fig:331sL12sL23} shows the allowed points in terms of $s_{eL12}^{}$ and $s_{eL23}^{}$ 
to simultaneously satisfy the neutrino oscillation data and the LFV constraints, 
except the red points which predict too large LFV rates.  
We see that that LFV constraints are not so stringent. 
The left panel of Fig.~\ref{fig:331-me1mueg} shows the allowed points to satisfy the current LFV bounds in terms of $m_{\eta_1^+}^{}$, 
  along with the future reach in Mu2e experiments~\cite{Bartoszek:2014mya} 
  at around ${\rm BR}(\mu\to e\gamma)= (2.5-6)\times 10^{-17}$ (horizontal lines). 
For the allowed point, we have also calculated the contribution to the muon $g-2$ from charged bosons. 
The left panel of Fig.~\ref{fig:331-me1damu} shows the predictions in terms of $m_{\eta_1^+}^{}$. 
We have found that the contribution is negative and cannot reconcile the discrepancy 
between the experimental result and the SM prediction. 
See \cite{Kelso:2014qka} for the contribution from extra gauged bosons, which is found to be positive.

\subsection{Inverted ordering}
In case of the Inverted ordering, {$(m_{\nu_1},m_{\nu_2},m_{\nu_3})=(m_{\nu_1},m_{\nu_2},0)$},
{
we have to fine-tune our free parameters to satisfy the neutrino oscillation data, since the neutrino mass matrix is nealy diagonal. Hence, the number of solution is very limited compared to the normal ordering case.  However, the LFV constraints would be automatically satisfied, once we find solutions. Such a situation is often seen in supersymmetric models (see, for example)~\cite{Mohapatra:2008wx}.
}
The input parameters vary in the following ranges, 
\begin{align}
&
33\ {\rm GeV}\le v_\eta\le 37\ {\rm GeV},
\ 
44\ {\rm GeV}\le v_\sigma\le 48\ {\rm GeV},
\ 
0.1\ {\rm GeV^2}\le (\delta m^{+2}_{\eta_1 h_1},\delta m^{+2}_{\eta_2 h_1}) \le 1\ {\rm GeV^2}
,\nn\\
&
100\ {\rm GeV}\le m_{\eta_{1,2}^+}\le1000\ {\rm GeV},\quad
-1\le s_{eL12}\le 0,\  0\le s_{eL23}\le 0.5,\ -1\le s_{eL13}\le -0.7
,\nn\\
&
-0.4\le s_{eR12}\le -0.1,\  0\le s_{eR23}\le 0.2,\ 0.4\le s_{eR13}\le 0.7.
\end{align}
{The mass parameters 
$m_{h_{1,2}^+}$, $\delta m^{+2}_{\eta_1 \eta_2}$, and $\delta m^{+2}_{h_1 h_2}$ can be written in terms of those above parameters,
and their mass scales are, respectively, $ m_{h_{1,2}^+}\approx {\cal O}$(100) GeV, and $\delta m^{+2}_{h_1 h_2} \approx \delta m^{+2}_{\eta_1 \eta_2}\approx  {\cal O}(1)\ {\rm GeV^2}$.
}
Then we show some representative figures of our results which simultaneously satisfy the neutrino oscillation data 
  and the constraints from the LFV processes. 
Here we have examined $2\times10^7$ sampling points to search for our allowed parameters. 
The right panel of Fig.~\ref{fig:331sL12sL23} shows the allowed points in terms of $s_{eL12}^{}$ and $s_{eL23}^{}$ 
to simultaneously satisfy the neutrino oscillation data and the LFV constraints. 
We see that that LFV constraints do not affect to our solutions as expected. 
The right panel of Fig.~\ref{fig:331-me1mueg} shows the allowed points to satisfy the current LFV bounds in terms of $m_{\eta_1^+}^{}$, along with the future reach in Mu2e experiments~\cite{Bartoszek:2014mya}
at around ${\rm BR}(\mu\to e\gamma)= (2.5-6)\times 10^{-17}$ (horizontal lines).
Comparing to the normal case, {the resulting ${\rm BR}(\mu\to e\gamma)$ is much smaller and}
more difficult to be detected even with the future experiments. 
For the allowed points, we have also calculated the contribution to the muon $g-2$ from charged bosons. 
The right panel of Fig.~\ref{fig:331-me1damu} shows the predictions in terms of $m_{\eta_1^+}^{}$. 
One finds that the result is similar to the case of normal ordering.

\begin{figure}[tbc]
\begin{center}
\includegraphics[scale=0.6]{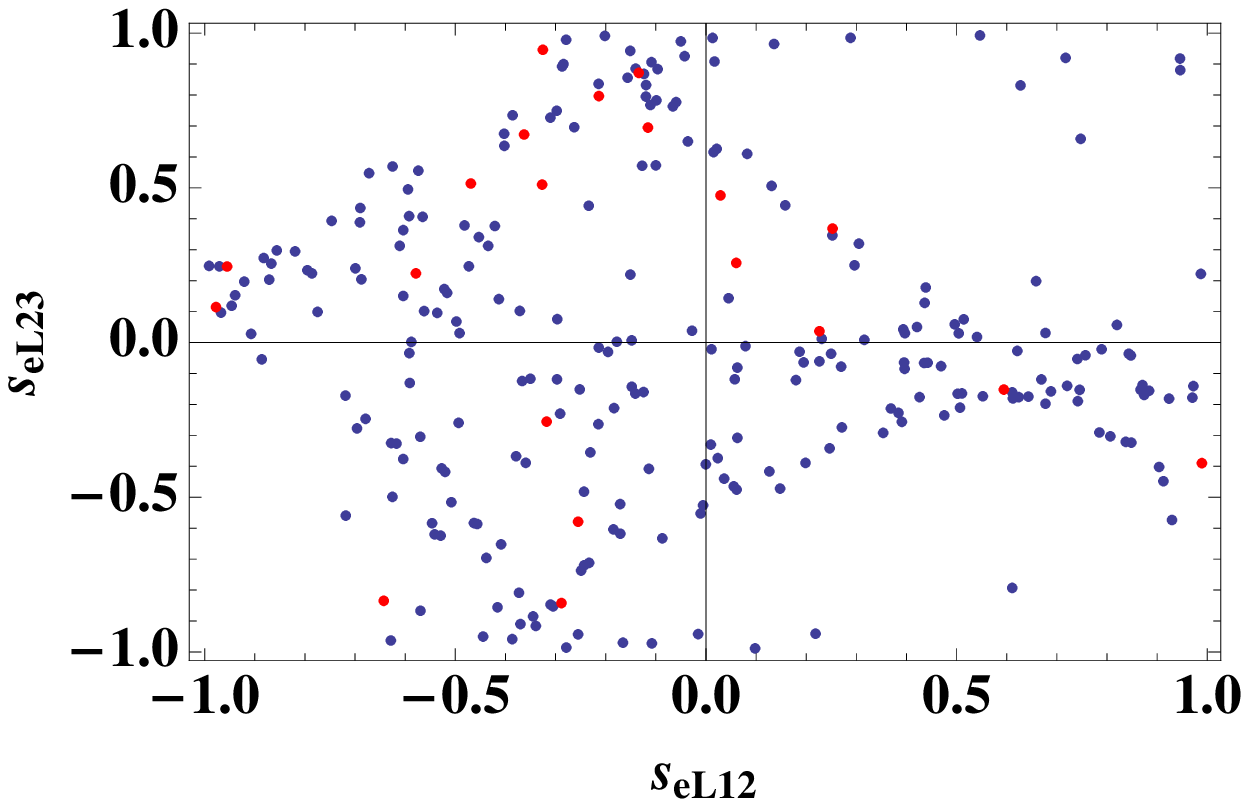}
\includegraphics[scale=0.6]{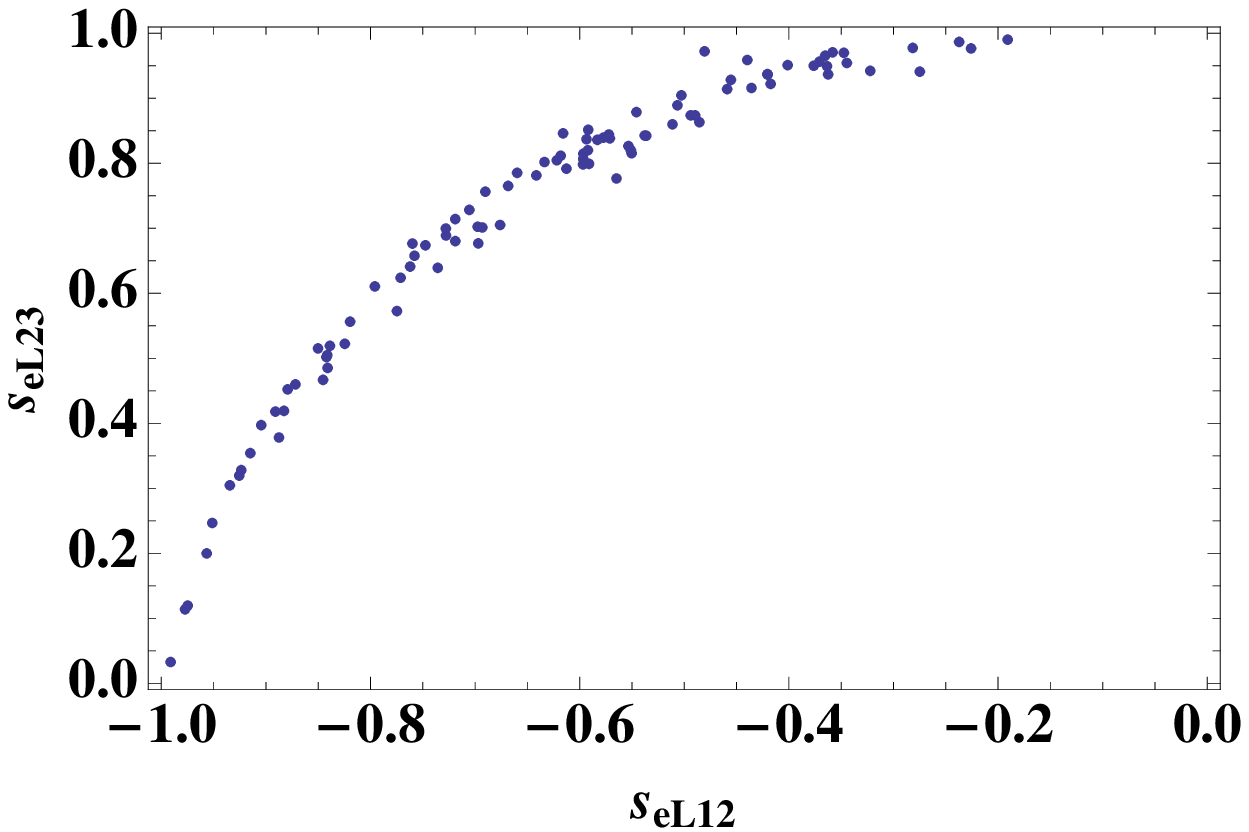}
\caption{Parameter scan in terms of $s_{eL12}$ and $s_{eL23}^{}$  to satisfy the LFV constraints and neutrino oscillation data. Here  the left panel is {for} the normal ordering case, {where} the red points do not satisfy the LFV constraints. The right panel is {for} the inverted ordering case.}
\label{fig:331sL12sL23}
\end{center}
\end{figure}

\begin{figure}[tbc]
\begin{center}
\includegraphics[scale=0.6]{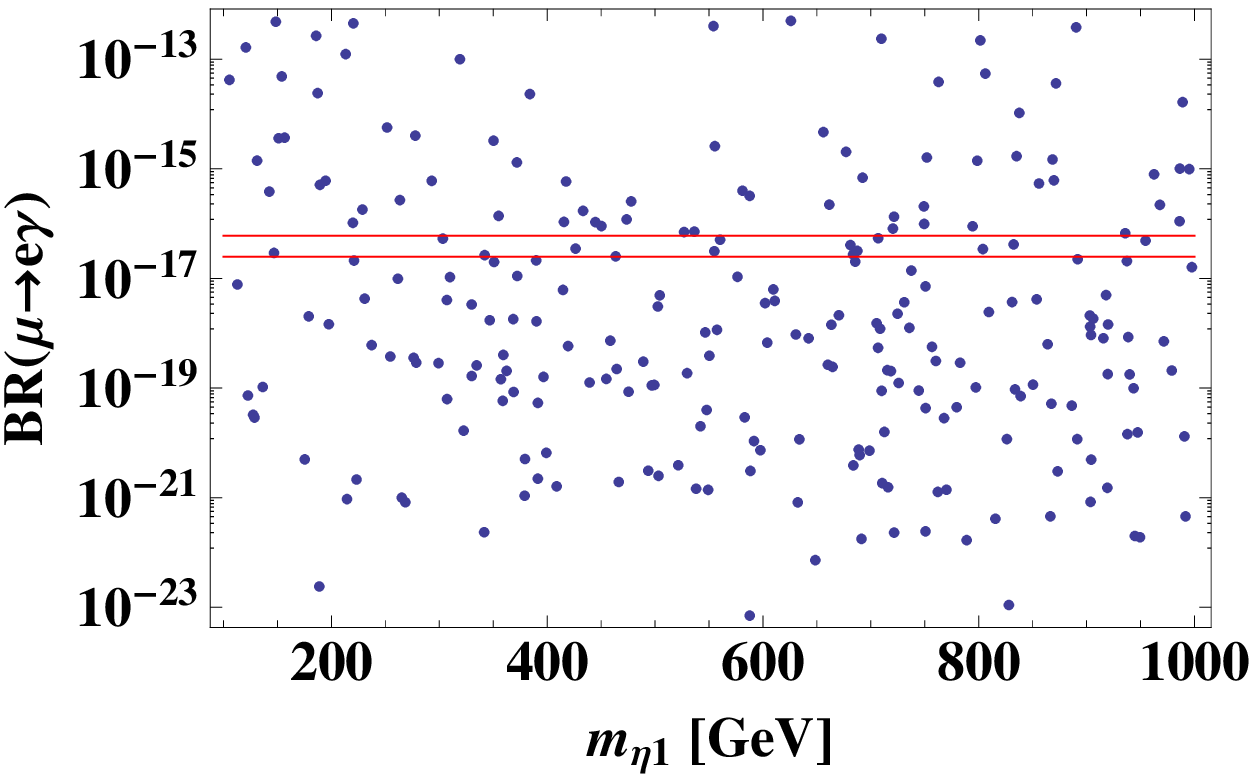}
\includegraphics[scale=0.6]{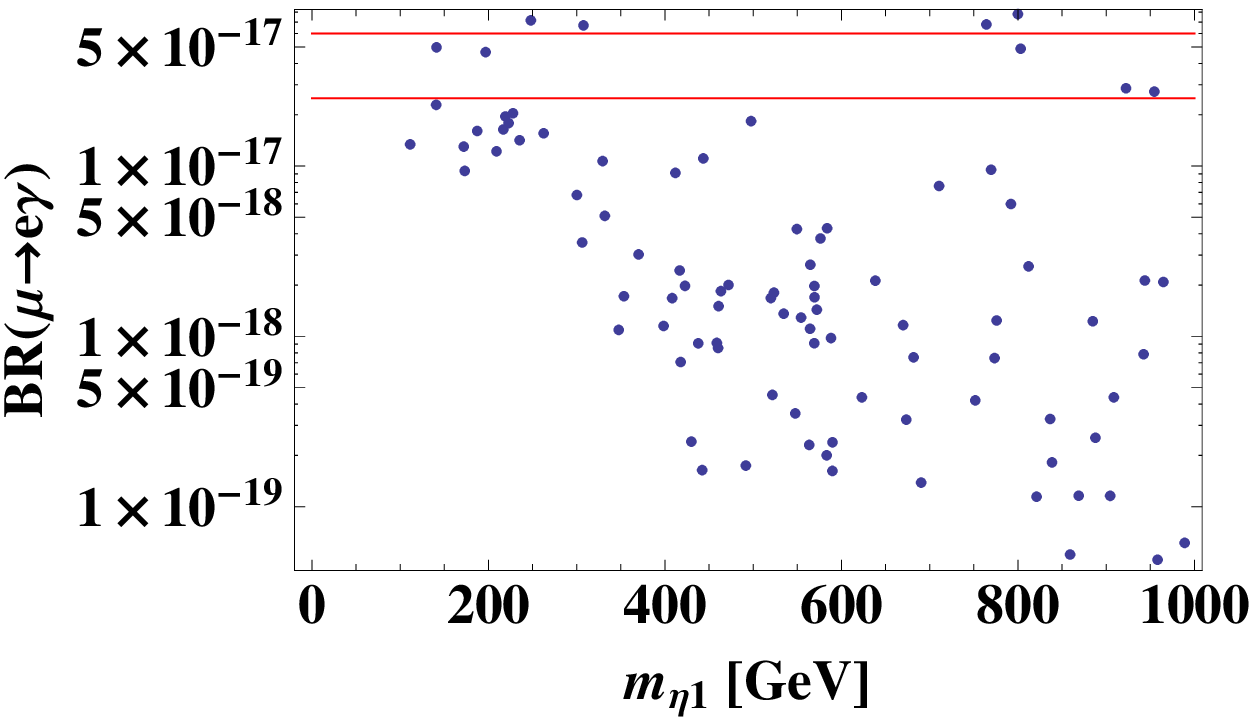}
\caption{Allowed points to satisfy the $\mu\to e\gamma$ constraint in terms of $m_{\eta_1^+}^{}$, where the horizontal lines denote the future reach by Mu2e experiments~\cite{Bartoszek:2014mya} at around ${\rm BR}(\mu\to e\gamma)= (2.5-6)\times 10^{-17}$. The left panel is {for} the normal ordering case, {while} the right panel is {for} the inverted ordering case.}
\label{fig:331-me1mueg}
\end{center}
\end{figure}

\begin{figure}[tbc]
\begin{center}
\includegraphics[scale=0.6]{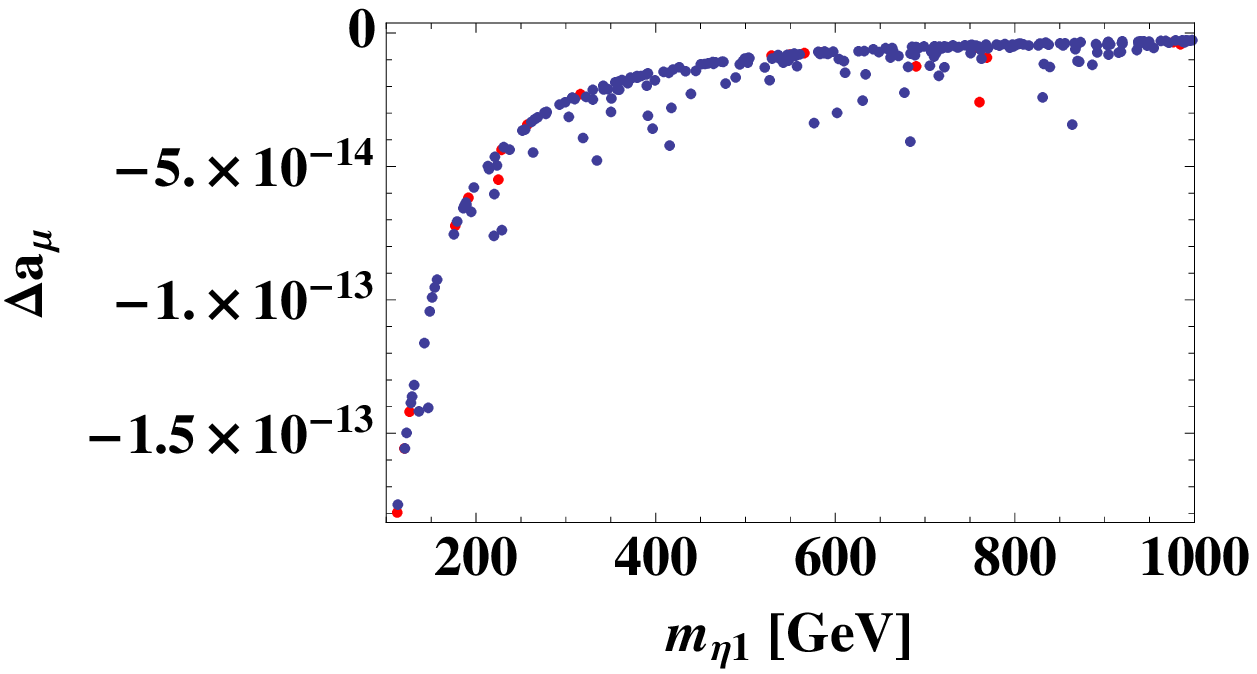}
\includegraphics[scale=0.6]{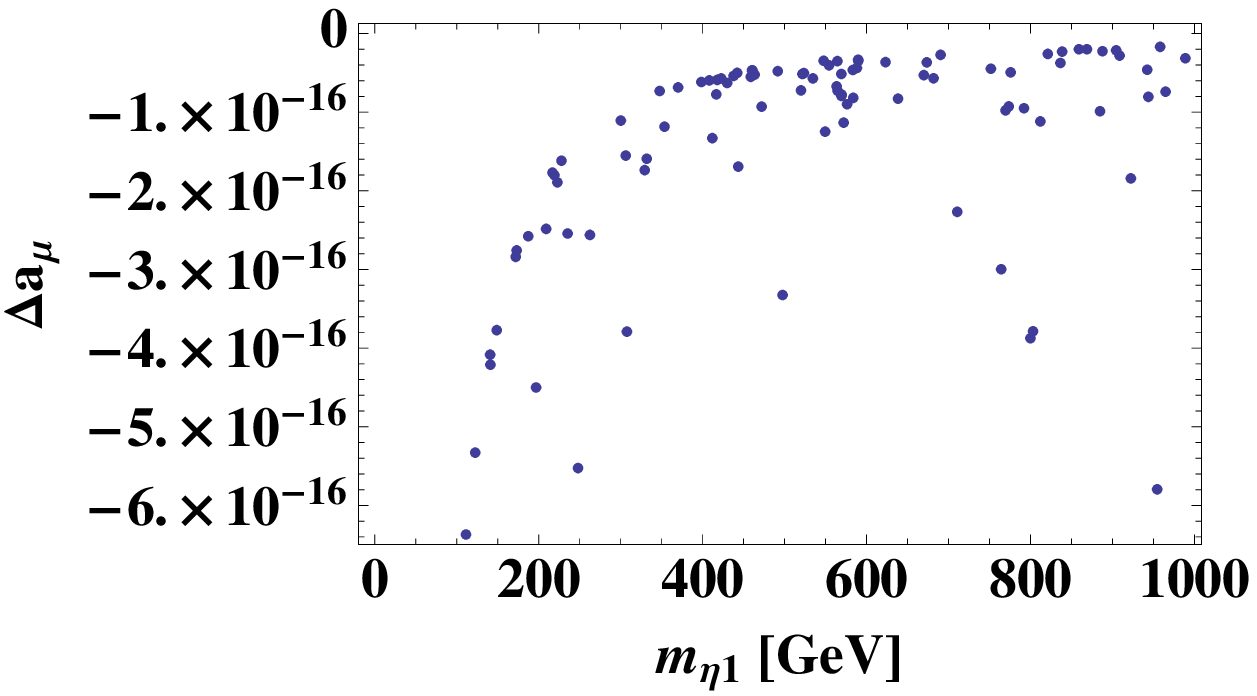}
\caption{Muon anomalous magnetic moment in terms of $m_{\eta_1^+}^{}$ to satisfy the LFV constraints and neutrino oscillation data, where the red points do not satisfy the LFV constraints. One can see that all the absolute values of data is less than $10^{-10}$.
Since the observed deviation from the SM prediction is positive and of ${\cal O}(10^{-9})$, the other contributions such as extra gauged bosons could be expected~\cite{Kelso:2014qka}.
Here the left panel is {for} the normal ordering case, {while} the right panel is {for} the inverted ordering case.}
\label{fig:331-me1damu}
\end{center}
\end{figure}


\section{Conclusions}
 
We have proposed a radiative seesaw model with a $SU(3)_C\times SU(3)_L\times U(1)_X$ gauge symmetry, 
in which the neutrino mass is induced through one-loop radiative corrections 
with the charged lepton mass. 
As a result, there is a strong correlation between the charged lepton and neutrino masses, 
and it is nontrivial if the current neutrino oscillation data are reproduced. 
In the model, the LFV processes are also induced via one-loop quantum corrections. 
We have performed general parameter scan for {the} normal and inverted {mass} ordering cases, and found that a large portion of parameter space 
can simultaneously satisfy the current neutrino oscillation data and the constraints on the LFV processes. 
The parameter region we have found can be partly tested by future Mu2e experiments. 
We have also calculated contributions to the muon anomalous magnetic moment 
from charged scalar particles in our model, and found that the contributions are not significant. 


\vspace{0.3cm}
{\it Acknowledgments}:
\noindent
This work was supported in part by the United States Department of Energy (N.O.), and 
  the Korea Neutrino Research Center which is established by the National Research Foundation of Korea (NRF) grant 
  funded by the Korea government (MSIP) (No. 2009-0083526) (Y.O.).

\end{document}